\documentclass{article}
\usepackage{url}
\usepackage{float}
\usepackage{comment}
\usepackage{setspace}
\usepackage{blindtext}
\usepackage{graphicx}
\usepackage{amsmath}
\usepackage{amsfonts}
\usepackage{lipsum}
\usepackage{wrapfig}
\usepackage{multirow}

\begin{document}
\title{\large\bf A Survey on the Detection of Android Malicious Apps\footnote{Reference: Springer, Advances in Computer Communication and Computational Sciences pp 437-446, 2019.}}
\author{Sanjay K. Sahay\footnote{BITS, Pilani, Dept. of CS \& IS, Goa Campus, Goa, India, Email: ssahay@goa.bits-pilani.ac.in} \hspace{0.15mm} and  Ashu Sharma\footnote{C3i Center, CSE, IIT Kanpur, India, Email: ashush@cse.iitk.ac.in}}

\date{}

\maketitle

\begin{abstract}
	Android-based smart devices are exponentially growing, and due to the ubiquity of the Internet, these devices are globally connected to the different devices/networks. Its popularity, attractive features, and mobility make malware creator to put number of 
malicious apps in the market to disrupt and annoy the victims. Although to identify the malicious apps, time-to-time  various techniques are proposed. However, it appears that malware developers are always ahead of the anti-malware group, and the proposed techniques by 
 the anti-malware groups are not sufficient  to counter the advanced malicious apps. Therefore, to understand the various techniques proposed/used for the identification of Android malicious apps, in this paper, we present a survey conducted by us on the work done by the researchers in this field.
\vspace*{0.1cm}
~
\end{abstract}

\section{Introduction}
 In 1992 International Business Machines Corporation came up with a prototype mobile computing device incorporating with the Personal Digital Assistant features (demonstrated  it in the Computers Dealer's Exhibition). Later on, Simon Personal Communicator first designed a device that was referred as smart device/phone which receives calls, sends faxes, emails and more. The smart devices/phone technology continued to advance throughout early 2000, and in 2007 Android-based smart device was unveiled by Google. Since then the popularity/demand for Android-based smart devices is continuously growing. An estimate shows that more than 15 billion smart devices are connected globe-wise and are expected to be reaching 200 billion by the end of the year 2020 \cite{macafeeq12017}. Also, due to the 
mobility and the attractive features of these devices have drastically changed 
our day-to-day life.  Many of these functionalities are very much similar to our other
information technology systems and are capable to remotely access the enterprise's data for the work. In addition, because of the ubiquity of Internet ubiquity, the user uses these devices for the shopping, financial transactions, share private information/data, etc. \cite{qr2015}. Hence, the security risks of the smart devices 
are now at never seen before levels, hence an attractive target for online attackers.
Also, nowadays online criminal are investing more and more for the sophisticated attacks, e.g., ransomware or to steal the valuable personal data/information from it. 

\par Today in the growing smart devices, Android  is the most popular operating system (OS) ($\sim$ 70\% of the market share) \cite{gandhewar2010google}, and are connected to different devices/networks through the Internet (five out of six mobile phones are Android based \cite{TRsym2016}). The popularity of Android OS is because of  its open source, java supported free rich software developer's kit,  exponential increase in the Android-based apps, and third-party distribution. According to  Statista,  in addition to the third-party Android apps, at Google Play store, there are $\sim$2 $\times10^6$ apps are available  for the users \cite{statista}, and some of these apps may be malicious \cite{9apps}. Therefore, the probability of the malware in the Android smart devices is now at never seen before levels. Thus attacks on the Android-based smart devices are increasing exponentially, basically due to the ease of creating the variants of malware \cite{vidas2011curbing} \cite{TR4}. In 2013, there was a 200\% increase in the malicious apps, and 3.7 million of variants added in McAfee's database \cite{Mcfee2015}. In 2015, Kaspersky reported that the growth rate of new malware variant is 300\% with 0.88 million new variants \cite{Kasper2015}. The number of malicious installations found in 2015 was around three million, and around seven thousand mobile banking Trojans were also found in the same year \cite{Kasper2014}.
In the 3rd quarter of 2015 Quick Heal Threat Research reported that per day they had received $\sim$$4.2 \times 10^5$ samples of the Windows and Android platforms \cite{qr2015}. Trend Micro estimated that the number of malicious mobile apps would reach 20 million by the end of 2017 \cite{trendmicro2016}. In this, stepping up its fight against bad apps and malicious developers, Google has removed over 700,000 Android apps (i.e., 70\% more apps that violated the Google Play policies in 2017 than the apps they removed in 2016) from Play Store and also took down as many as 100,000 bad developers in 2017 \cite{google2018}.

\par The recent attacks on Android devices show that the security features in these devices are not good enough to stop the adversary \cite{sym2017}. Therefore, its a need of time to design a robust anti-malware, in particular, to counter zero-day attack \cite{sharma2014evolution}. Also, it is an open question, how to detect the variants of malicious Android apps which are concealed in the 3rd party apps markets \cite{9apps}, and how to find the repackaged apps in the ocean of Android apps. In addition, to avoid the deployed detection methods, malware developer uses various obfuscation techniques \cite{TRsym2014}. Hence, any gap in the security of the Android smart devices will allow the attacker to access the information stored in it. To defend the 
attack/threat from the Android malicious apps, a number of static and dynamic methods were proposed \cite{arp2014drebin}, \cite{narayanan2016adaptive},\cite{rashidi2017android}, \cite{feizollah2017androdialysis}. 
But still, it appears that to defend the malicious apps, the proposed techniques are not good enough in the growing smart devices usage in our day-to-day life \cite{sharma2014evolution}. Thus, Android based smart devices security is one of the important fields to be 
addressed, and understanding the market share of the Android-based smart devices, in this paper to know the various techniques proposed/used to identify the Android malicious apps, in section 2, 
to understand security mechanism of the Android devices we discuss in brief that how user data and its resources are secure by providing the features viz.  sandbox, permission, secure inter-component communication,  and signing the apps. In section 3, we present the survey conducted by us on the work done by the researchers in this field. Finally, in section 4, we summarize our conclusion.

\section{Security Mechanisms}
The security of the Android-based devices are mainly focused to protect the user data, its resources, and isolation of the application by providing the features viz.  sandbox, permission, secure inter-component communication,  and signing the apps. In this, 
the {\bf sandbox} in the Android based system isolate the apps from each other by the user ID (UID) and permissions.  After installation, the apps runs in its assigned sandbox and can access only its own resources, unless other apps give explicit access permission to this applications. However, if the apps are designed by the same developer, then such apps can share the same UID and can run in the same sandbox to share resources/data between them.

\par Android apps consist of four components viz., services, broadcasts, activities, and providers. Similar to Inter Processes Communication in the Linux system, it provides a secure
{\bf Inter-Component Communication} by binder mechanism (middleware layer of
Android).  Inter-Component Communication is achieved by the intents/message, and these intents are explicitly used for the communication, if it identifies the receiver name, or used for implicit communication that allows the receiver to know that
can it access this intent or not.

\par {\bf Application signing} creates a certification
between developers and their applications to ensure the security of the apps, and before putting it in its sandbox, it makes a relationship between the apps and UID. 
Without application signing, apps will not run. If two or more apps have the same UID, then
all the apps which have the same UID can communicate with each other, share the permissions and can run in the same 
sandbox.  By the application singing, apps update process can be simplified. The updated new version will have all the permissions that the old version has, and also the certificate does not change so that the package manager can verify the certificate.  It also makes sure that without using Inter-Component Communication, apps cannot communicate with another apps. 
However, if the apps are developed by the same developer, then the developer without changing the 
application signing  can enable the direct communication  between the same developer apps.

\par In these smart devices, there are four levels of {\bf permissions} viz. signatureOrSystem (granted to the apps that are installed by
the root or pre-installed apps), signature (granted within the same sandbox), dangerous (granted by the
users) and normal (granted automatically) permissions. In total there are 235 permissions out of which 163 are hardware accessible and remaining are for user information access \cite{olmstead2016apps}. Before installation of the apps, the system asks the users to grant all requested permissions, if users agree and grant the requested permissions to the apps, then in general installation becomes successful else it may get canceled. These permissions mechanisms put some restriction when the apps want to access the application programming interface (API) which are sensitive to the OS. The Android apps run in the sandbox, and if it needs to access resources/data outside its sandbox which could potentially affect the user's privacy/data viz. {\it short message service}, contacts, camera, location, etc. then the user has to approve/reject the permission.

\section{Detection of Android Malicious Apps}

\par To identify the Android malicious apps,  static and dynamic analysis are the two basic methods that are used \cite{kapratwar2016static}. In both the methods classifiers are first trained with a dataset for the classification of apps. 
However, in static analysis, apps are analyzed without executing it to extract some patterns viz. APIs used, data flow, permissions, control flow, intents, etc. Whereas, in the dynamic analysis the codes are analyzed during the execution of the apps, and monitoring 
its dynamic behavior (resources usage, tracing system calls, API call, network traffic, etc.) and the response of the system.  In this, in 2009, understanding that the users do not understand what applications will do with their data/resources, and thus not able to decide which permissions shall be allowed to the application to run with,  Fuchs, et al.,  \cite{fuchs2009scandroid} proposed a tool called SCANDROID (suppose to be the first program analysis tool for the Android-based devices) which can extracts security specifications from the applications, and can identify that the data flow of such apps are consistent with the specification or not.

\par In 2012 Sanz. et al. \cite{sanz2012automatic}  based on machine-learning techniques proposed a method to detect the Android malicious apps by automatically characterize the applications. For the classification, the feature sets used are the printable strings, apps permissions, and the apps permissions extracted from the Android Market. Their experiment with seven different categories (820 samples) and five classifiers shows that  among the selected five classifiers, Bayes Tree Augmented Naive Bayes is the best classifier (0.93 area under the curve (AUC)), while random forest (RF) stands second in the investigated classifier (0.9 AUC), and among the analyzed classifier, the worst was Decision Tree with J48 (0.64 AUC). Wu. et al., based on the static feature proposed a technique called  \textit{DroidMat}, to detect the Android malicious apps which analyze AndroidManifest.xml and the systems calls. For the experiment, they used 238 malicious and 15000 benign programs and claimed that their approach is very effective (97.87\% accuracy), scalable and efficient \cite{wu2012droidmat}.

\par In 2013 Michael et al. proposed a mobile-sandbox to automatically analyze the Android apps in two steps. In the first step (static analysis), applications Manifest files are parsed and decompiled; then they find that the applications are using suspicious permissions/intents or not. In the next step, dynamic analysis is performed, where the apps are executed to know all the actions 
including those originating  from the associated API calls. They experimented with $\sim$ 36,000 apps from the third-party Asian mobile markets and reported that 24\% of all analyzed apps use associated calls. \cite{spreitzenbarth2013mobile}. Min Zheng et al. \cite{zheng2013droid} developed a signature based system called {\it DroidAnalytics} for collecting the malware automatically, and to 
generate signatures for the identification of the malicious code segment. They conducted extensive experiments with 150,368 apps and detected 2,494 malicious apps from one hundred two different families, in which three hundred forty-two of them were zero-day malicious samples from the six different families. They claimed that their methods have significant advantages over the traditional MD5 hash based signature, and can be a valuable tool for the classification of Android malicious apps.

\par In 2014, a detection method was proposed by Quentin et al. which depends on the opcode-sequences. They tested libsvm and SVM classifier with the reduced data set (11,960 malware and 12,905 benign applications) and obtained 0.89\% F-measure. However, their approach is not capable to detect completely different malware \cite{jerome2014using}. Kevin Allix et al. \cite{allix2014large} devised classifiers that depend on the features set that are designed from the apps control flow graphs. They analyzed their techniques with $\sim$50,000 Android apps and claimed that their approach outperformed existing machine learning approaches. Also from the analysis, they concluded that for the realistic malware detectors, the 10-fold cross-validation approach on the usual dataset is not a reliable indicator of the performance of the classifier.

\par The smartphone can act as like a zombie device, controlled by the hackers via command and control servers. It has been found that mobile malware are targeting Android devices to get root level access so that from the remote server they can execute the instructions. Hence, such type of malware will be a big threat to the homeland security. Therefore Seo, et al. \cite{seo2014detecting} discusses the defining characteristics which are inherent in the devices and shown the feasible mobile attack scenario against the Homeland security. They analyze various mobile malware samples viz. monitoring the home and office, banking, flight booking and tracking, from both the unofficial and official market to identify the potential vulnerabilities. 
Their analysis shows that the two banking apps (axis and mellat bank app) charges SMS for malicious activities and two other banking apps was modified to get permissions without the consent. Finally, they discuss an approach that mitigates the Homeland Security from the malware threats.

\par In 2015, Jehyun Lee et al. developed a technique to detect the malicious apps that use automated extraction of the family signature. They claimed that compare to earlier behavior analysis techniques their proposed family
signature matching  detection accuracy is high and can detect variants of known
malware more accurately and efficiently than the legacy signature matching. Their results were based on the analysis done with the 5846 real Android malicious apps which belong to 48 families collected in April 2014 and achieved 97\% accuracy. Smita et al. 
\cite{naval2015employing} addressed the problem of system-call injection attack (inject independent and irrelevant system-calls when programs are executing) and proposed a solution which is evasion-proof and is not vulnerable to the system-call injection attacks.
 Their technique characterizes the program semantics by using the property of asymptotic equipartition, which allows to find the information-rich call sequences to detect the malicious apps. According to their analysis, the semantically-relevant paths can be applied to know the malicious behavior and also to identify the number of unseen/new malware. They claimed that  the proposed solution is robust against the system-call injection attacks and are effective to detect the real malware instances.

\par In 2016, a host-based Android malicious apps detection system called Multi-Level Anomaly Detector for Android Malware ({\it MADAM}) has been proposed by Saracino, et al. \cite{saracino2016madam}. Their system at the same time can analyze and correlates the 
features at 4 levels viz., application, kernel, user and package to identify.  It stopped the malicious behaviors of one hundred twenty-five known malware families, encompasses most of the malware. They claimed that MADAM could understand the behaviors 
characteristics of almost all the real malicious apps which can be known in the wild, and it can block more than 96\% of 
Android malicious apps. Their analysis on 9,804 clean apps shows low false alarm rate, limited battery consumption (4\% energy overhead), and negligible performance overhead. BooJoong et al. \cite{kang2016n}, proposed an n-opcode  based static analysis for the  classification and categorizing the Android malware. Their approach does not utilize the defined features viz. permissions, API 
calls, intents, and other application properties, rather it automatically discovers the  features that eliminate the need of an expert to find the required promising features. Empirically they showed  that by using the occurrence of n-opcodes, a reasonable classification accuracy can be achieved, and for n = 3 and n = 4, they have achieved F-measure up to 98\% for categorization and classification of the malware.

\par Based on inter-component communication (ICC) related features, Ke Xu et al. \cite{xu2016iccdetector} proposed a method to identify the malicious apps called ICCDetector, which can capture the interaction between the components or cross application boundaries. They evaluated the performance of their approach with 5264 malware and 12026 benign apps and achieved an accuracy of 97.4\%. 
Also, after manually analyzing, they discovered 43 new malware in the benign data and reduced the number of false positive to seven. Jae-wook Jang et al. \cite{jang2016andro} proposed a feature-rich hybrid anti-malware system called {\it Andro-Dumpsys}, which can 
identify and classify the malware groups of similar behaviour. They claimed that {\it Andro-Dumpsys} could detect the malware and classify the malware families with low false positive (FP) and false negative (FN). It is also scalable and capable to respond zero-day threats. Gerardo Canfora et al. \cite{canfora2016hmm} evaluated a couple of techniques for detecting the malicious apps. First one was based on {\it hidden markov model}, and the 2nd exploits the structural entropy. They claimed that their approach is effective for Desktops viruses, and can also classify the malicious 
apps. Experimentally they achieved a precision of 0.96 to differentiate the malicious apps, and 0.978 to detect the malware family \cite{canfora2016hmm}. For the detection of malware in runtime, 
Sanjeev Das et al. \cite{das2016semantics} proposed a hardware-enhanced architecture called{\it GuardOL}  by using {\it field programmable gate arrays} and processor. Their approach after extracting the system calls made the features from  the high-level semantics of the malicious behavior. Then the features are used to  train the classifier and multilayer perceptron to detect the malicious apps during the execution. The importance of their design was that the approach in the first 30\% of the execution detects 46\% of the malware, and after 100\% of execution 97\% of the samples have been identified with 3\% FP. \cite{das2016semantics}.

\par In 2017, Ali Feizollah et al.  \cite{feizollah2017androdialysis} proposed AndroDialysis to  evaluate how effective is the Android intents (implicit and explicit) as a feature to identify the malicious apps. 
They showed that the intents are semantically rich features compared to well studied other features viz. permissions, to know the intentions of malware. However, they also concluded that such features are not the ultimate solution, and it should be used together with other known promising features. Their result was based on the analysis of the dataset of 7406 apps (1846 clean and 5560 infected apps). They achieved a 91\% detection accuracy by 
 using Android Intent, 83\% using Android permission and by combining both the features they obtained the detection rate of 95.5\%. They claimed that for the malware detection, Intent is more effective than the permission. 
Bahman Rashidi et al. \cite{rashidi2017android} proposed an Android resources usage risk assessment called {\it XDroid}. They claimed that the malware detection accuracy could be increased significantly by using the temporal behavior tracking, and security alerts of the suspicious behaviors can be generated. Also, in real-time their model can inform  users about the risk level of the apps, and can dynamically update the parameters of the model by the user's preferences and from the on-line algorithm. They conducted the experiment on benchmark {\it Drebin} malware dataset and demonstrated that their approach could estimate the risk levels of the malware up to  $82\%$ accuracy and with the user input it can provide an adaptive risk assessment.

\par Recently, Ashu et al. \cite{sharmasahay} with the benchmark {\it Drebin} dataset and first without making any groups, they examine the five classifiers using the opcodes occurrence as a feature for the identification of malicious applications and achieved detection
 accuracy up to 79.27\% by functional tree classifier. They observed that the overall accuracy is mainly affected by the FP. However, highest true positive (99.91\%) is obtained
 by RF and fluctuates least with the number of 
features compared to the remaining four classifiers. Their analysis shows that overall accuracy mainly depends on the FP of the investigated classifiers. Later on, in 2018, similar to the analysis of Windows Desktop malware classification \cite{sharma2016effective} \cite{sharma2016improving} \cite{sahay2016grouping}, they group-wise analyzed the dataset based on permissions and achieved up to 97.15\% overall average accuracy 
\cite{AshuIJNS}. They observed that the MICROPHONE group detection accuracy is least while CALENDAR group apps are detected with maximum accuracy, and top 80-100 features are good enough to achieve the best accuracy. As far as the true positive (TP) is concerned, RF gives best TP for the CALENDAR group.

\section{Summary}

\par The attack/threat by the malware to the Android-based smart devices which are connected to different devices/networks through the Internet is increasing day-by-day. Consequently, these smart devices are highly vulnerable to the advanced malware, and its effect will be highly destructive if an effective and timely 
counter-measures are not deployed. Therefore, time-to-time, various static and dynamic methods that have been proposed by the authors for the identification of the Android malicious apps have been discussed in this paper to counter the advanced malicious apps in the fast-growing Internet and smart devices usage into our daily life. Although in literature various survey viz. \cite{parvez2015survey} \cite{kimberly2017survey} \cite{souri2018state} are available on the 
detection  Android malicious  but in this survey, we presented the comparative study of various approaches and the observations done by the various authors understanding that all malware are not of the same type and cannot be detected with one algorithm.

\bibliographystyle{plain}
\bibliography{references}

\end{document}